# A Simulation Approach to Multi-station Solar Irradiance Data Considering Temporal Correlations

Xingbo Fu, Feng Gao, Jiang Wu
System Engineering Institute
Xi'an Jiaotong University
Xi'an, China
{xbfu, fgao, jwu}@sei.xjtu.edu.cn

Ruanming Huang, Yichao Huang, Fei Fei
State Grid Shanghai Electric Economic and Technology Research Institute
Shanghai, China
hrmjobhrms@163.com, {huangyc, fief}@sgcc.com.cn

*Abstract*—Solar energy is one of important renewable energy sources and simulation of solar irradiance can be used as input for simulation of photovoltaic (PV) generation. This paper proposes a simulation algorithm of multi-station solar irradiance data considering temporal correlations. First of all, we group all the days of the observed data to k clusters for each station based on their daily features of solar irradiance and the daily states constitute Markov chain of days. Then, we reduce state permutations of different stations before getting Markov Transition Probability Matrix (MTPM). In terms of the observed data and MTPM, the simulation approach is proposed. Finally, we test our approach by applying to solar irradiance data of three stations and show that the properties of simulated data match those of the observed data.

## I. INTRODUCTION

Renewable energy sources, especially solar and wind, have drawn tremendous attentions from researchers. As for solar energy, solar irradiance is one of the most significant factors that influence PV power generation. Solar irradiance, however, is uncertain and intermittent due to the erratic meteorological conditions including cloud amount, clearness, dust and relative humidity. These make analyses of solar irradiance complicated.

Numerous sensors have been deployed to monitor the environment and record the meteorological data. Because of some geo-sensor glitches, missing weather data may compromise the performance of power system analysis such as PV generation prediction. Simulation of solar irradiance data centers on modeling solar irradiance time series and simulating new data close to historical data for planning and operation of power system.

Some researchers have shown their attempts to deal with solar irradiance problems. C. W. Richardson put forward stochastic simulation of solar radiation by using a multivariate model [1]. A.P. Grantham et al. presented a method to generate synthetic sequences of daily and hourly global horizontal irradiation by developing a model to deal with the deterministic component of global horizontal irradiation, and then adding a stochastic component using a nonparametric bootstrapping technique [2]. V. Prema et al. showed trend pattern and seasonal pattern of solar irradiance while predicting solar irradiance data [3]. Mellit et al. proposed a model for generating sequences of global solar radiation data for isolated sites by using artificial neural network and Markov transition matrices (MTM) [4].

We presented a simulation method of solar irradiance data of one station based on feature clustering and Markov chain in our early works [5]. This method may be used to generate long samples of solar radiation and it can deal with the situation where we simulate one-station solar irradiance data. On the other hand, when we apply this model to multi-station data, each station is considered as an individual site regardless of their correlations among each other.

Solar irradiance data from National Renewable Energy Laboratory (NREL) include one-minute-interval global horizontal irradiance (GHI) and diffuse horizontal irradiance data at several solar radiation monitoring stations [6]. And in this paper, we concentrate on GHI—total hemispheric shortwave irradiance as measured by an Kipp & Zonen Model with calibration factor traceable to the World Radiometric Reference (WRR) [7]. Table I shows details of three stations and Fig. 1 shows GHI data of these stations from June 17[th], 2015 to June 21st, 2015.

Table I  Details of Three Stations

| Station | Latitude | Longitude | Location |
|---|---|---|---|
| Solar Radiation Research Laboratory (SRRL) | 39.74° North | 105.18° West | Golden, Colorado |
| Solar Technology Acceleration Center (STAC) | 39.76° North | 104.62° West | Aurora, Colorado |
| Oak Ridge National Laboratory (ORNL) | 35.93° North | 84.31° West | Oak Ridge, Tennessee |

The research presented in this paper is supported in part by the State Grid Science and Technology Program of China.

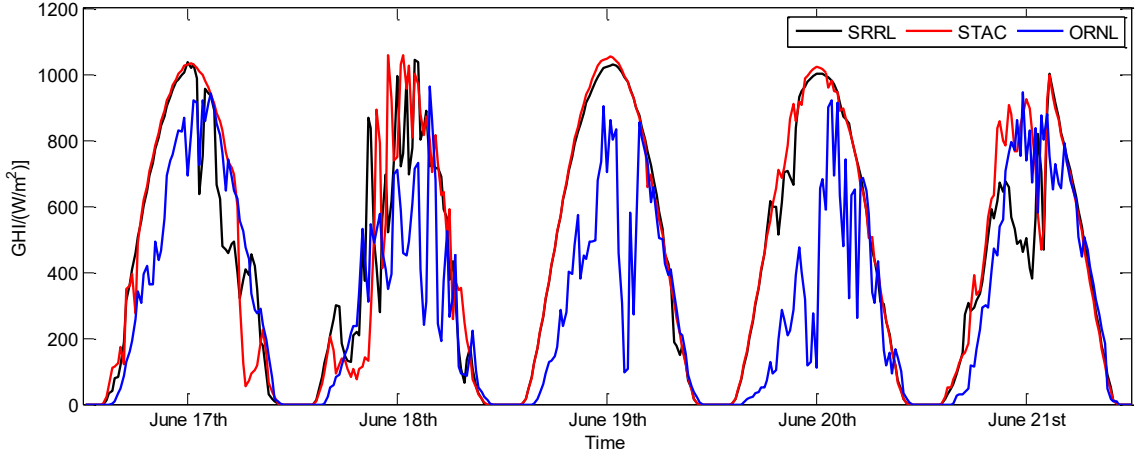

Figure 1 GHI Data of Three Stations (June 17th, 2015~ June 21st, 2015)

In order to simulate multi-station GHI data, we propose a simulation approach in this paper and the main contributions are as follows:

- Considering correlations of stations, we construct Markov chain of multi-station daily states;
- State space reduction is applied before getting MTPM in order to reduce dimensions of MTPM;
- We optimize the simulation process to improve the performance.

The rest of this paper is as follows. Firstly, we introduce the simulation approach in section II. Next, the effectiveness of the proposed approach is tested in section III by using GHI data of three stations. Finally, conclusions will be provided in section IV.

## II. SIMULATION APPROACH TO MULTI-STATION GHI DATA

### A. Daily Features of GHI Data

As a kind of time series, one-minute-interval GHI data of one site consists of 1,440 daily samples. Considering intermittency, we keep 1,080 daily samples from 3:00 a.m. to 9:00 p.m. as the observed GHI data. Obviously, these samples are too large to be inputs for some algorithms like clustering so it is necessary to extract daily features of GHI data.

Some statistical values can be used to characterize a time series [8]. For instance, the standard deviation can quantify the amount of dispersion of a time series while skewness is a measure of the asymmetry of the probability distribution of a time series about its mean. Besides, the fluctuation intensity of GHI is also a significant feature in this paper. To address this problem, we introduce Reverse Fluctuation Count (RFC), Average Fluctuation Magnitude (AFM) and Moving Fluctuation Intensity (MFI) [9] then we can characterize GHI fluctuations by MFI.

RFC is defined as count value that when fluctuation trend is reversed, then count value add 1. Fluctuation trend is reversed when

$$[x(i+1) - x(i)][x(i) - x(i-1)] < 0. \qquad (1)$$

AFM is defined as

$$AFM = \frac{1}{N}\sum_{i=1}^{N} abs[x(i+1) - x(i)]. \qquad (2)$$

And MFI is defined as

$$MFI = RFC \times AFM \qquad (3)$$

Table II lists five features that we select as daily features of GHI data.

Table II List of Daily Features

| Feature Name | Description |
|---|---|
| Mean | $\bar{x} = \frac{1}{N}\sum_{i=1}^{N} x(i)$ |
| Standard Deviation | $\sigma = \sqrt{\frac{1}{N-1}\sum_{i=1}^{N}(x(i)-\bar{x})^2}$ |
| Skewness | $\gamma = \frac{1}{N}\sum_{i=1}^{N}\left(\frac{x(i)-\bar{x}}{\sigma}\right)^3$ |
| Kurtosis | $\beta = \frac{1}{N}\sum_{i=1}^{N}\left(\frac{x(i)-\bar{x}}{\sigma}\right)^4 - 3$ |
| Moving Fluctuation Intensity | $MFI = RFC \times AFM$ |

So these five daily features make up feature vector $f$ of each day. And for station $i$, feature vectors $f_i$ of $n$-day historical GHI data are as follows:

$$f_i = (f_i^1, f_i^2, \dots, f_i^m \dots, f_i^{n-1}, f_i^n) \qquad (4)$$

## B. Clustering Algorithm

Clustering is a statistical technique of grouping objects in clusters and objects belonging to the same cluster are as similar as possible to each other. Typical cluster models include connectivity-based clustering, distribution-based clustering, centroid-based clustering and density-based clustering [10]. In this paper, we adopt k-means clustering and group all the days to $k$ clusters $C = \{c_1, c_2, \ldots, c_k\}$ based on their normalized feature vectors of GHI data. In general, the appropriate $k$ for GHI data is 4 [11].

In order to figure out differences in the $k$ clusters, the GHI curves of all the centroids are drawn and according to these curves each cluster is defined as a state of PV generation suitability. That is to say, the smooth and high-mean curve is more suitable for generation while the erratic and low-mean one is not.

Therefore, we get states $s^i$ of $n$-day historical GHI data for the station $i$ as follows:

$$s^i = (s_1^i, s_2^i, \ldots, s_t^i, \ldots, s_n^i) \quad (5)$$

and states matrix $s$ of all these $j$ stations are as follows:

$$s = \begin{pmatrix} s^1 \\ s^2 \\ \vdots \\ s^i \\ \vdots \\ s^j \end{pmatrix} = \begin{pmatrix} s_1^1, s_2^1, \ldots, s_t^1, \ldots, s_n^1 \\ s_1^2, s_2^2, \ldots, s_t^2, \ldots, s_n^2 \\ \vdots \\ s_1^i, s_2^i, \ldots, s_t^i, \ldots, s_n^i \\ \vdots \\ s_1^j, s_2^j, \ldots, s_t^j, \ldots, s_n^j \end{pmatrix} \quad (6)$$

$$s_t^i \in C, i = 1,2,\ldots,j, t = 1,2,\ldots,n \quad (7)$$

And on day $t$, states $s_t$ of the whole $j$ stations are as follows:

$$s_t = \begin{pmatrix} s_t^1 \\ s_t^2 \\ \vdots \\ s_t^i \\ \vdots \\ s_t^j \end{pmatrix} \quad (8)$$

So we get states matrix

$$s = (s_1, s_2, \ldots, s_t, \ldots, s_n) \quad (9)$$

## C. State Space Reduction

In (6), the $t$ th column of states matrix $s$ stands for states of the whole $j$ stations on day $t$. When we group days to $k$ clusters, there are $k^j$ possible state permutations. For example, we will get $4^3 = 64$ state permutations with 4 clusters and 3 stations. However, we notice that not all these state permutations happen in the real world. Hence, state space reduction can be applied before calculating MTPM so that the dimensions of MTPM can be reduced. It means that if the state permutation $s_t$ does not happen in $n$-day historical GHI data, we will cut off this state permutation from state space. When $r_0$ different state permutations do not happen, the size of reduced state permutations are

$$r = k^j - r_0 \quad (10)$$

We convert this reduced state space into digital code in Table III.

Table III  Digital Code and Reduced State Space

| Digital Code | Reduced State Space |
|---|---|
| 1 | $(c_1, c_1, \ldots, c_1)$ |
| 2 | $(c_1, c_1, \ldots, c_2)$ |
| $\vdots$ | $\vdots$ |
| $r-1$ | $(c_k, c_k, \ldots, c_{k-1})$ |
| $r$ | $(c_k, c_k, \ldots, c_k)$ |

Consequently, state matrix $s$ can be wrote as follows:

$$s = (s_1, s_2, \ldots, s_t, \ldots, s_n)$$
$$\to s' = (d_1, d_2, \ldots, d_t, \ldots, d_n) \quad (11)$$
$$d_t = 1,2,\ldots,r$$

## D. MTPM

A Markov chain is a type of Markov process that has a discrete state space [12]. This stochastic process that satisfies the Markov property - the future state $d_{t+1}$ depends only on the present state $d_t$ and it does not depend upon the previous state $d_1, d_2, \ldots, d_{t-1}$ [13]. We can formulate this property in mathematical notation as follows:

$$P\{s'(t+1) = d_{t+1}|s'(1) = d_1, s'(2) = d_2, \ldots, s'(t) = d_t\} \quad (12)$$
$$= P\{s'(t+1) = d_{t+1}|s'(t) = d_t\}.$$

In order to describe the changes of states, Markov chain $\{d_1, d_2, \ldots, d_t, \ldots, d_n\}$ can be characterized by the first-order MTPM $P$ consisting of the transition probabilities

$$P = \begin{pmatrix} p_{11} & \cdots & p_{1r} \\ \vdots & \ddots & \vdots \\ p_{r1} & \cdots & p_{rr} \end{pmatrix} \quad (13)$$

$$p_{ij} = P\{s'(t+1) = j|s'(t) = i\} \quad (14)$$
$$for\ i,j = 1,2,\ldots,r$$

## E. Simulation Approach

This part presents the simulation approach to multi-station GHI data in terms of MTPM and the observed GHI data. The approach is as follows:

Step 1: Set the initial random daily state $s_{init}$ and size of days to simulate $n$. $s' = \{s_{init}\}$.

Step 2: Given a random number $r \in (0,1)$, find out the next daily state $s_{next}$, which satisfies

$$p_{s_{init}\ s_{next}} < r < \sum_{i=1}^{s_{next}} p_{s_{init}\ i} \quad (15)$$

Step 3: Append $s_{next}$ to $s'$ and let $s_{init} = s_{next}$.

Step 4: When the size of elements in $s'$ reaches $n$, go to Step 5. Otherwise, return to Step 2.

Step 5: Convert $s'$ into state matrix $s$ according to Table III.

Step 6: As for $s_t^i$ in $s$, select one day whose state is $s_t^i$ from the observed days.

Step 7: Connect the raw data of days from Step 6 to get the simulated data.

### III. CASE STUDY

In this section, the performance of this simulation approach is presented based on GHI data of three stations: SRRL, STAC and ORNL.

#### A. GHI Data

We collect the observed one-minute-interval GHI data from January 1$^{st}$, 2012 to December 31$^{st}$, 2018. Table IV shows the Pearson correlation coefficients between each two stations among them [14].

Table IV Pearson correlation coefficients

| Stations | Pearson correlation coefficients |
|---|---|
| (SRRL,STAC) | 0.908 |
| (SRRL,ORNL) | 0.697 |
| (STAC,ORNL) | 0.732 |

As Table IV shows, SRRL and STAC have a higher correlation because of their closer location.

Considering different sunshine durations among different seasons, we divide days of the observed GHI data into four seasons (Table V).

Table V Size of Days

| Seasons | Size of Days |
|---|---|
| Spring | 625 |
| Summer | 644 |
| Autumn | 644 |
| Winter | 644 |

#### B. Results of Clustering

We group GHI data shown in Part A into four clusters for different seasons at SRRL, STAC and ORNL stations. Fig. 2, Fig. 3 and Fig. 4 shows curves of average one-minute-interval GHI data of all the days from the same cluster for each season at these three stations.

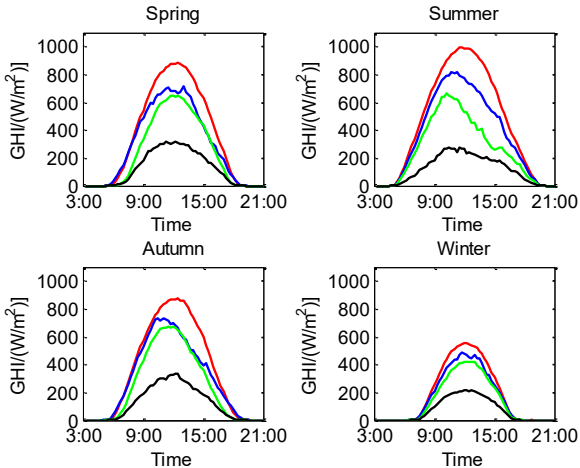

Figure 2 Average SRRL's GHI Data from the same cluster

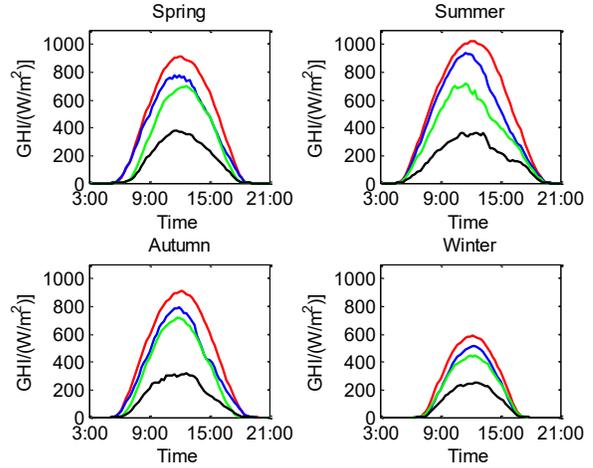

Figure 3 Average STAC's GHI Data from the same cluster

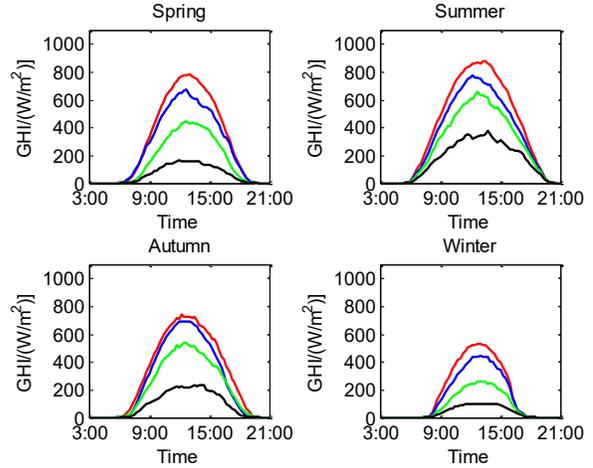

Figure 4 Average ORNL's GHI Data from the same cluster

As shown in Fig. 2, Fig. 3 and Fig. 4, the red curves are the most suitable for generation while the black the worst. Therefore, the red curves represent $c_1$, the blue curves represent $c_2$, the green curves represent $c_3$ and the black curves represent $c_4$.

#### C. Result of Simulation

We simulate one-year GHI data of these three stations and Fig. 5 shows the curve of simulated GHI data from July 5$^{th}$ to July 9$^{th}$.

#### D. Performance of Simulation

##### 1) Statistical Analysis

In this part, we compare some statistical properties of the simulated one-year GHI data with those of the observed GHI data. Table VI shows the mean and standard deviation (Std) of the observed GHI data and the simulated data each day.

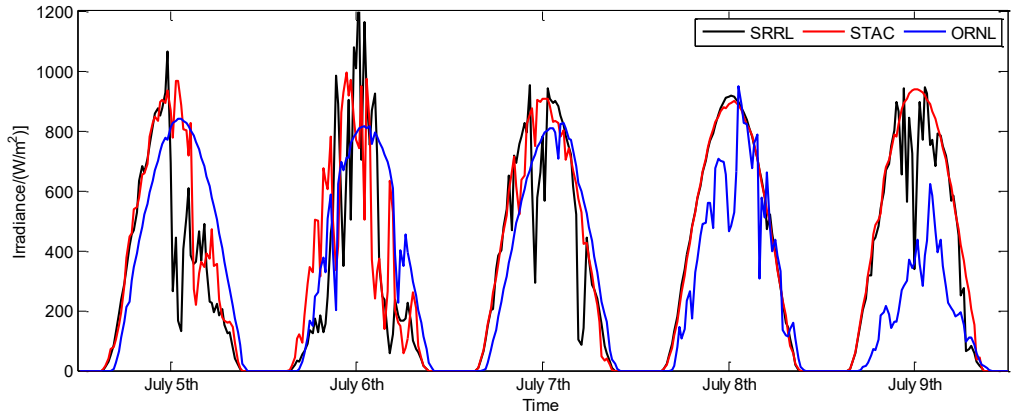

Figure 5 Simulated GHI Data from July 5th to July 9th

Table VI Statistical Properties of Simulated Data and Observed Data

| Year | SRRL Mean | SRRL Std | STAC Mean | STAC Std | ORNL Mean | ORNL Std |
|---|---|---|---|---|---|---|
| 2012 | 265.63 | 270.56 | 284.01 | 285.43 | 226.39 | 233.43 |
| 2013 | 261.12 | 268.44 | 278.28 | 282.39 | 208.68 | 219.92 |
| 2014 | 253.42 | 262.10 | 270.03 | 276.04 | 220.99 | 229.38 |
| 2015 | 249.86 | 259.60 | 270.13 | 276.96 | 212.61 | 219.72 |
| 2016 | 265.30 | 273.03 | 282.77 | 285.26 | 227.32 | 233.60 |
| 2017 | 254.76 | 263.61 | 268.42 | 274.58 | 206.31 | 213.24 |
| 2018 | 260.36 | 267.30 | 272.40 | 276.65 | 193.79 | 201.04 |
| All | 258.64 | 266.38 | 275.16 | 279.62 | 213.74 | 221.49 |
| Simulation | 260.76 | 266.15 | 281.26 | 282.34 | 211.27 | 221.61 |

*2) Cumulative Distribution*

Cumulative distribution function (CDF) describes the cumulative probability associated with a random variable and the approximate CDFs between simulated GHI data and the observed data is required [15]. Fig. 6 shows CDFs of these three station consisting the observed GHI data and the simulated GHI data.

*3) Monthly Curves*

We present monthly curves of the observed GHI data every year and the simulated data for each station shown as Fig. 7, Fig. 8 and Fig. 9.

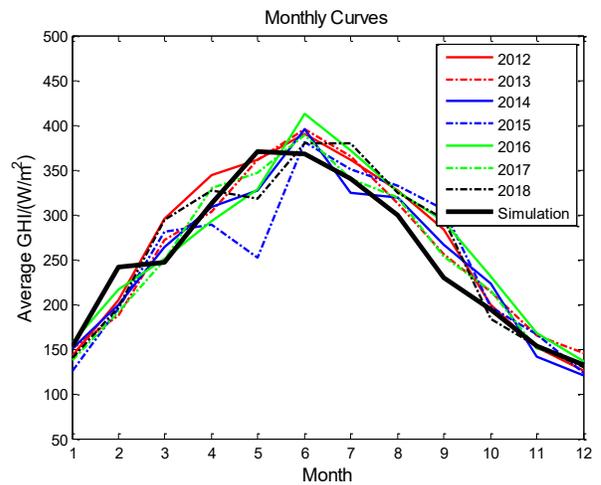

Figure 7 SRRL Monthly Curves

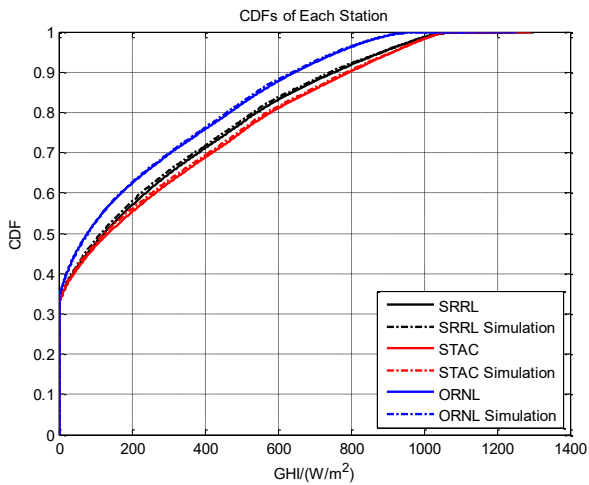

Figure 6 CDFs of Simulated Data and the Observed Data

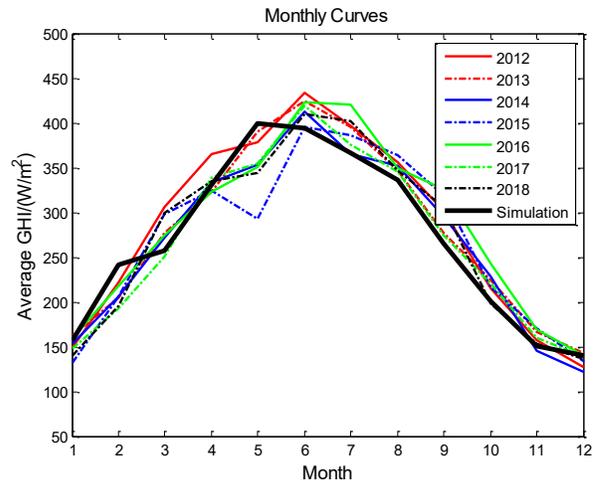

Figure 8 STAC Monthly Curves

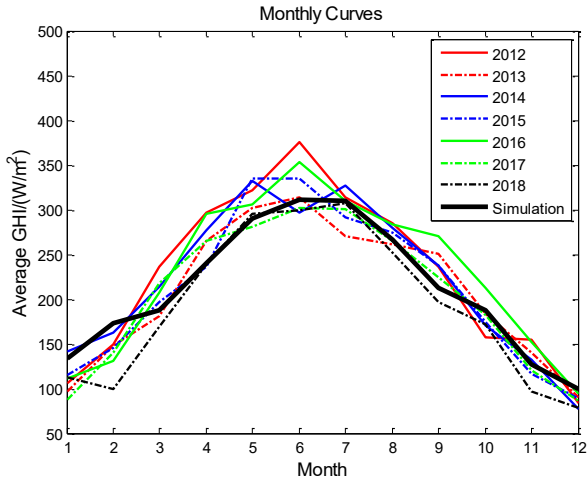

Figure 9 ORNL Monthly Curves

## IV. Conclusion

A simulation approach to multi-station solar irradiance data considering correlation is presented in this paper. We extract five features of daily GHI data and group days of the observed solar irradiance data into four clusters. Each cluster represents one state of solar power generation suitability and daily states of the observed data constitute Markov chain. Before getting MTPM, state space reduction can reduce the dimensions of MTPM. We propose the simulation approach base on MTPM and the observed GHI data. Finally, estimations with three stations are given.

In our future work, we may concentrate on simulation of large-scale stations and deal with curse of dimensionality. Besides, feature extraction of daily data can also be another improvement.

## References

[1] Richardson C W. Stochastic simulation of daily precipitation, temperature, and solar radiation[J]. Water Resources Research, 1981, 17(1):182–190.
[2] Grantham A P, Pudney P J, Boland J W. Generating synthetic sequences of global horizontal irradiation[J]. Solar Energy, 2018, 162: 500-509.
[3] Prema V, Rao K U. Development of statistical time series models for solar power prediction[J]. Renewable energy, 2015, 83: 100-109.
[4] Mellit A, Benghanem M, Arab A H, et al. A simplified model for generating sequences of global solar radiation data for isolated sites: Using artificial neural network and a library of Markov transition matrices approach[J]. Solar Energy, 2005, 79(5): 469-482.
[5] Fu X, Gao F, Wu J, et al. A Simulation Method of Solar Irradiance Data Based on Feature Clustering and Markov Transition Probability Matrix[C]//2018 13th World Congress on Intelligent Control and Automation (WCICA). IEEE, 2018: 1741-1746.
[6] https://midcdmz.nrel.gov
[7] https://midcdmz.nrel.gov/apps/go2url.pl?site=STAC
[8] Brillinger D R. Time series: data analysis and theory[M]. Siam, 1981.
[9] Zhang W, Liu Z. Simulation and analysis of the power output fluctuation of photovoltaic modules based on NREL one-minute irradiance data[C]//2013 International Conference on Materials for Renewable Energy and Environment. IEEE, 2014, 1: 21-25.
[10] Jain A K, Murty M N, Flynn P J. Data clustering: a review[J]. ACM computing surveys (CSUR), 1999, 31(3): 264-323.
[11] Li S, Ma H, Li W. Typical solar radiation year construction using k-means clustering and discrete-time Markov chain[J]. Applied Energy, 2017, 205: 720-731.
[12] Song Z, Geng X, Kusiak A, et al. Mining Markov chain transition matrix from wind speed time series data[J]. Expert Systems with Applications, 2011, 38(8): 10229-10239.
[13] Barsotti F, De Castro Y, Espinasse T, et al. Estimating the transition matrix of a Markov chain observed at random times[J]. Statistics & Probability Letters, 2014, 94: 98-105.
[14] Benesty J, Chen J, Huang Y, et al. Pearson correlation coefficient[M]//Noise reduction in speech processing. Springer, Berlin, Heidelberg, 2009: 1-4.
[15] Miller F P, Vandome A F, Mcbrewster J. Cumulative Distribution Function[J]. 2006.